# Dynamic Stimulation of Superconductivity With Resonant Terahertz Ultrasonic Waves

Alan M. Kadin, *Senior Member, IEEE* and Steven B. Kaplan, *Senior Member, IEEE*

*Abstract*—An experiment is proposed to stimulate a superconducting thin film with terahertz (THz) acoustic waves, which is a regime not previously tested. For a thin film on a piezoelectric substrate, this can be achieved by coupling the substrate to a tunable coherent THz electromagnetic source. Suggested materials for initial tests are a niobium film on a quartz substrate, with a BSCCO intrinsic Josephson junction (IJJ) stack. This will create acoustic standing waves on the nm scale in the thin film. A properly tuned standing wave will enable electron diffraction across the Fermi surface, leading to electron localization perpendicular to the substrate. This is expected to reduce the effective dimensionality, and enhance the tendency for superconducting order parallel to the substrate, even well above the superconducting critical temperature. This enhancement can be observed by measuring the in-plane critical current and the perpendicular tunneling gap.  A similar experiment may be carried out for a cuprate thin film, although the conduction electrons might be more responsive to spin waves than to acoustic waves.  These experiments address a novel regime of large momentum transfer to the electrons, which should be quite distinct from the more traditional regime of large energy transfer obtained from direct electromagnetic stimulation. The experiments are also motivated in part by novel theories of the superconducting state involving dynamic charge-density waves and spin-density waves.  Potential device applications are discussed.

## I. Introduction

NONEQUILIBRIUM STIMULATION of superconductivity has a long history as a subject of research [1]-[3], but practical applications have been limited, in part because effects have generally been quite small.  Early work focused on raising the energy of excitations of the superconducting state.  While pumping energy into a superconductor is often equivalent to heating, in certain cases the nonequilibrium electron distribution may correspond to cooling the electrons [4], [5]. This was demonstrated for a variety of excitations, including microwaves, GHz-frequency phonons, optical radiation, and electron tunneling. Observed effects included enhanced critical current, energy gap, and in some cases small increases in critical temperature $T_c$ on the mK scale [6]-[8]. Some recent work [9], [10] has evaluated the effect of THz electromagnetic radiation on superconducting films.  Furthermore, many researchers have recently used ultrafast laser pulses and observed the transient nonequilibrium optical response of cuprates and other high temperature superconductors [11]-[15].

In contrast to this earlier approach to nonequilibrium superconductivity that coupled large energies into the electrons, we are proposing here to couple large changes in *momentum*, sufficient to diffract electron waves from one side of the Fermi surface to the other. We note that a momentum change $\Delta p = \Delta E/u$ for both photons and phonons, but because phonons have a speed $u$ that is a million times slower than the speed of light, large momentum transfer from a phonon is more easily obtained.  Such large momentum changes are also associated with elastic scattering from impurities and boundaries, and also with diffraction from parallel planes of modulated charge density.  Such diffracting planes are present in Bloch waves, and similar planes are self-induced in charge-density waves (CDW), also known as Peierls distortions [16], [17]. In both cases, the plane spacing (typically on the nm scale) corresponds to a wavevector

$$\mathbf{q} = 2\mathbf{k}_F + \mathbf{G} \qquad (1)$$

where $\mathbf{k}_F$ is the electron wavevector at the Fermi surface, and $\mathbf{G}$ is a reciprocal lattice vector.  Such diffracting planes lead to electron localization and an energy gap in the electron energy spectrum.

But do CDWs have anything to do with superconductors? Even before the BCS theory, Fröhlich [18] suggested that the CDW diffracting planes and localized electrons could move together without resistance, forming the basis for superconductivity, at least in 1D.  But when real CDW materials were discovered, they were found to be insulators. Evidently, the moving CDWs pin on impurities, preventing free motion of the combined structure [17].  While the CDW superconductor theory was largely forgotten after BCS, it is worth noting that both theories incorporate much of the same electron-phonon formalism, with even the same equations for the energy gap and $T_c$ [16].

The diffracting planes in Bloch waves and CDWs are static, but the wavevector in Eq. (1) may alternatively correspond to a dynamic phonon mode, at a high frequency typically on the THz scale.  Such trans-Fermi-surface phonons are known to interact with conduction electrons, as initially derived by Kohn [19]. More recently, Kadin and Kaplan [20]-[23] have suggested that a three-dimensional dynamic CDW based on self-induced standing waves of these phonons might be responsible for superconductivity in some cases. This would be similar to the proposal of Fröhlich, except that the high frequency of the dynamic CDW might prevent pinning, thus permitting true supercurrents.

There is little direct experimental evidence for such a dynamic CDW in conventional superconductors, but the neutron scattering results of Aynajian [24] suggested a possible dynamic density wave at $f \sim 2\Delta(0)/h \sim 0.7$ THz, for both Nb and Pb. These results have been analyzed by several groups [25]-[27], but the interpretation remains somewhat unclear. There is also some evidence relating CDWs or similar spin-density waves (SDW) in cuprates and other high-temperature superconductors [28]-[30], although the significance for superconductivity remains controversial. In addition, evidence has been observed for transient optical enhancement of superconductivity related to dynamic phonons or CDWs, and several theories have been proposed to explain these results [31]-[34]. This remains a subject of widespread current interest.

In the present paper, we propose application of external coherent phonons to a superconducting film, over a range of frequencies that should include at least one set of Kohn phonons with wavevector given by Eq. (1), in order to generate ultrasonic standing-waves that might interact with electrons so as to induce or enhance superconductivity. Such an experiment should be feasible using modern device technology, as described below. A much earlier paper [35] proposed application of similar THz phonons to induce a CDW state, but this was not pursued further. Recent experiments have observed enhancement of CDW and SDW states under pulsed laser illumination [36], [37], which may be related.

## II. PROPOSED EXPERIMENT

The concept of the proposed experiment is shown in Fig. 1. This shows a flat piezoelectric substrate, coated with superconducting films on both sides. A THz voltage source is connected to the two electrodes, creating a uniform THz electric field vertically across the substrate. This, in turn, generates a vertical acoustic vibration in the piezoelectric substrate [38]. Given partial reflection and partial transmission at planar interfaces, this should yield acoustic standing waves with nodes parallel to the interfaces in both the substrate and the films. We are targeting a frequency ~ 0.5-1 THz, comparable to the gap frequency of Nb. The corresponding wavelength is a few nm, so that the surfaces of the piezoelectric substrate must be flat to near atomic smoothness, in order to avoid scattering of the direction of the acoustic wave at the surface [39]. We suggest using Nb or Pb for the initial experiments. These are cubic materials with fairly low anisotropy, so that the usual polycrystalline films may be acceptable.

Fig. 1 also shows two probes of the superconducting film. First, the electrical characteristics parallel to the substrate may be measured using a patterned narrow strip, including the critical current $I_c$ in the superconducting state, the resistance in the normal state, and the critical temperature $T_c$ defining the boundary between them. Second, the electrical properties perpendicular to the film can be measured using a vertical tunnel junction. This should permit the energy gap $\Delta(T)$ to be measured, and also the critical current of a Josephson junction. The tunnel junction may be a standard Nb/AlOx/Nb tunnel junction used for Nb Josephson junctions.

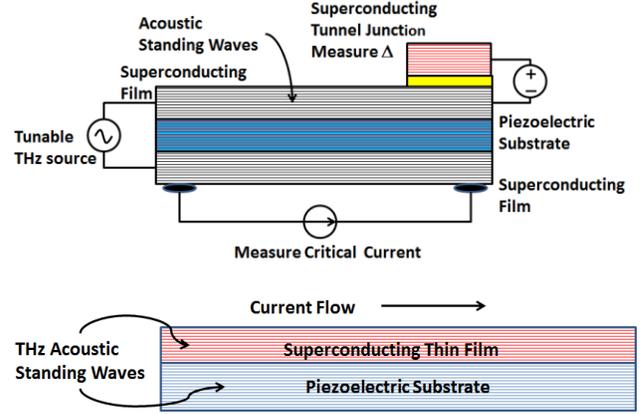

Fig. 1. Conceptual Schematic Diagram of proposed experiment. (a) A tunable THz source applies an ac voltage across a piezoelectric substrate, which generates acoustic waves in both the substrate and in superconducting films deposited on the substrate. The critical current and the energy gap of the superconducting film are measured as a function of frequency, amplitude, and temperature. (b) Enlarged region showing the nodal planes of the acoustic standing waves parallel to substrate, and current flowing parallel to the planes.

An equally important aspect of the experiment is the tunable THz source, and how the THz signal is coupled to the piezoelectric substrate. A leading candidate for a tunable THz source is an intrinsic Josephson junction (IJJ) stack, consisting of a single crystal of the cuprate BSCCO [40]-[42]. Fig. 2 shows one possible configuration for coupling the THz radiation into a piezoelectric substrate, which may be a quartz single crystal. The BSCCO crystal is mounted on the side of the quartz substrate, with common superconducting Nb electrodes for both crystals, forming a stripline resonator. The characteristic impedance of the stripline can be selected to enable efficient coupling of the THz radiation. A terminating resistance at the far end of the stripline resonator may be used to avoid reflection of the THz radiation back toward the source.

An IJJ stack consists of a series array of $N$ nominally identical Josephson junctions, where $N$ may be of order thousands. Ideally, all of these operate at the same frequency, so that by varying the total dc voltage $V = Nhf/2e$, one can smoothly vary the frequency, over a range as large as $0.3 - 2.4$ THz. The target range for the proposed experiment may be $0.5 - 1.0$ THz, so this would appear to be sufficient. However, while BSCCO IJJs can function over a range of temperatures below about 70 K, some parts of the frequency range may be accessible only at certain temperatures. So having the superconducting film under test at the same temperature as the IJJ may limit the tuning flexibility.

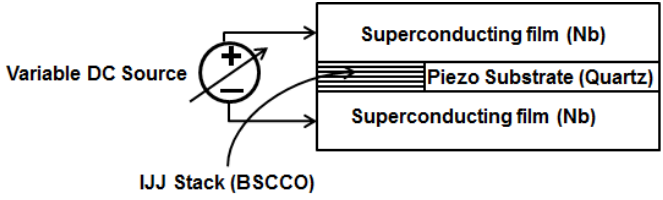

Fig. 2. Configuration of experiment with IJJ stack as tunable THz source. The IJJ is mounted on the side of the piezoelectric substrate, with common superconducting electrodes coupling the THz radiation into the piezoelectric.

Furthermore, self-heating in the IJJ may be significant, possibly heating the Nb film as well if they are mounted together. For this reason, there may be advantages to mounting the IJJ separately from the piezo device, with separate temperature control. Of course, this may also make the radiation coupling more difficult, possibly requiring quasi-optical coupling with lenses and antennas. So the simple configuration of Fig. 2 may be a reasonable first attempt, but may not produce optimum results.

A further control issue is varying the amplitude of the IJJ THz signal coupled to the piezo substrate. Since one usually does not have the ability to bias separate parts of the stack, it is difficult to vary the THz power. The total power provided by an IJJ can be up to ~ 1 mW, which is likely to be sufficient. However, the optimum power level is not well known, particularly given uncertainties in piezoelectric conversion efficiency, acoustic transmission coefficients, and electron-phonon interactions.

### III. ANTICIPATED EXPERIMENTAL RESULTS

The proposed experiment lies in a regime that has not yet been explored, where the standard theory may be incomplete. Therefore, any prediction of results is necessarily approximate and even speculative. But we can describe qualitatively what we expect, focusing on changes in the critical current $I_c$ and the energy gap parameter $\Delta$ of the superconducting film, as shown in Fig. 3.

We propose to examine the dependence of $I_c$ and $\Delta$ as a function of frequency and temperature. At the optimum frequency, we are looking for an enhancement in both $I_c$ and $\Delta$, both below and above the equilibrium $T_c$, as suggested in Fig. 3. $I_c$ probes the in-plane superconducting order, while $\Delta$ corresponds to current perpendicular to the film.

Overall, we anticipate that resonant acoustic standing waves corresponding to Eq. (1) may yield electron localization perpendicular to the film, leading to an increased tendency toward superconductivity in-plane. In contrast, the non-resonant response is likely to be a broad suppression, rather than an enhancement. Some suppression effects may be related to substrate heating or electron heating

The axes in Fig. 3 do not show the size of the anticipated effect, and indeed, we can only speculate. However, if superconductivity is indeed related to dynamic CDWs [20], [21], then we would anticipate a strong response, with perhaps up to a factor of two enhancement in the critical temperature.

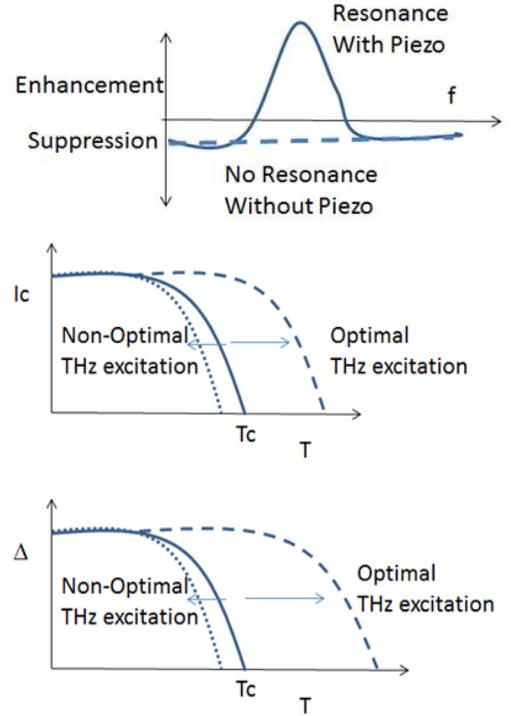

Fig. 3. Qualitative expected results from experiment of Figs. 1 and 2. (a) Enhancement of $I_c$ and $\Delta$ at resonant frequency, but weak suppression at other frequencies. (b) From in-plane transport measurement, enhancement of $I_c$ for optimal THz excitation, including increase in effective $T_c$. (c) From perpendicular tunneling measurement, enhancement of superconducting energy gap, including increase in effective $T_c$.

The same probes can be used to measure other parameters as well. For example, an SIS Josephson junction detector can be used to observe a Shapiro step due to the THz source at $V = nhf/2e$, thus calibrating the THz emission [43]. It may also show photon-assisted tunneling steps associated with the gap structure, at $V = 2\Delta/e \pm hf/e$. In the normal state above $T_c$, the resistance in the horizontal channel may show changes due to electron localization.

Another useful calibration would be to substitute a non-piezoelectric substrate in Fig. 1. This might permit one to distinguish any direct electromagnetic THz effects from those associated with resonant phonons.

Consider the real-space picture electrons and acoustic waves in Fig. 4, following the alternative model of [21]. This shows electrons that are both localized and correlated. The horizontal planes are externally imposed and are fixed in space. These should localize the electrons vertically (with larger acoustic amplitude creating greater compression), but the electrons bound to these horizontal planes will also be fixed in space, corresponding more to an insulator than a superconductor. The effective reduction from three to two dimensions is expected to be similar to an anisotropic 2D superconductor such as the cuprates and other high-$T_c$ materials. This may increase the tendency to form an induced dynamic CDW with vertical planes which that can move freely in the horizontal direction, thus enabling a superconducting state at an enhanced temperature.

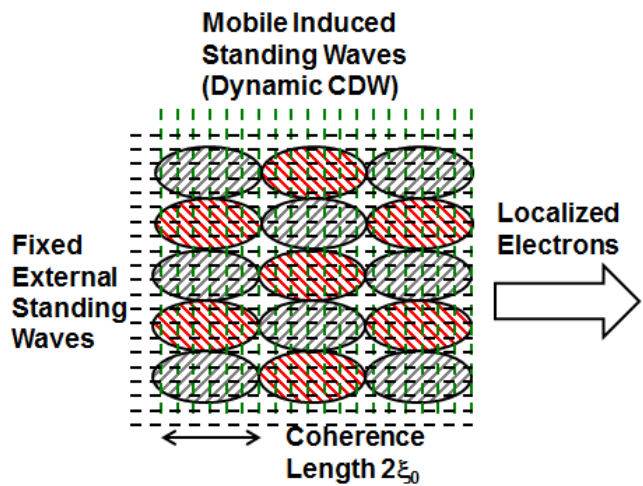

Fig. 4.  Real-space picture of correlated electrons in an anisotropic superconductor [21], with fixed externally imposed horizontal standing wave planes, and mobile induced vertical standing wave planes.  The localized electrons are free to move with the vertical planes in the horizontal direction.

Note that the configuration of Fig. 4 requires that the acoustic standing waves are fully in the horizontal plane, which in turn requires that the interfaces and surfaces are all smooth on the scale of the acoustic wavelength (a few nm). Any rough surfaces will give rise to diffuse reflection and transmission [39], leading to fixed planes in arbitrary directions, and restricting the flow of supercurrent.  In that case, one might expect to measure an energy gap, but without evidence of a zero-voltage current.  This would be more like a classic static CDW, where pinning prevents true superconductivity.

## IV.  Discussion

We have suggested a single resonant frequency, but there may be multiple frequencies for which Eq. (1) is satisfied, particularly for a polycrystalline film with an anisotropic Fermi surface.  If so, this may provide significant insight into the structure of the Fermi surface.

We have suggested initial tests using Nb thin films, but other materials are also possible.  Pb is known to have strong electron-phonon coupling, so one would expect Pb to show a large enhancement effect due to external phonon pumping. But it would also be of interest to test materials that have weak electron-phonon coupling, such as Al, Cu, and Au.  Al has an very low $T_c$ = 1.2 K, while Cu and Au never become superconducting.  It would be of interest to see whether superconductivity could be induced in these materials, perhaps with a relatively large acoustic perturbation.

There are also a variety of high-temperature superconductors, including $MgB_2$, cuprates such as YBaCuO, and pnictides such as GdFeAsO, and even $H_2S$ under pressure. Some of these may be associated with phonon-induced superconductivity, while others (such as the cuprates and pnictides) may be more related to spin waves.  Once this stimulation technique becomes established with simple superconductors, it would be of interest to do similar tests for these more exotic superconductors. A further extension would be to use a magnetoelectric substrate to generate standing spin waves [44], which may also permit diffraction across the Fermi surface if Eq. (1) is obeyed.  Superconductivity in some of these materials may be related to dynamic spin-density waves (SDWs), rather than CDWs [30].

It may also be useful to investigate alternative piezoelectric substrates [45], which have not been fully characterized in the THz regime.  For example, many perovskite ferroelectric materials, such as $SrTiO_3$, are also piezoelectric at low-temperatures [46].  These may be also be more compatible than quartz for testing the cuprates and other exotic materials.

Finally, if the proposed nonequilibrium enhancement of superconductivity can be demonstrated, there may be a variety of possible device applications. For example, this may provide a fast transient response, which relaxes quickly after the stimulation is turned on or off.  This might enable a fast switch or memory cell.

## V.  Conclusion

In summary, we have proposed an experiment to investigate a novel type of nonequilibrium effect in superconductors. This involves the generation of dynamic acoustic standing waves at THz frequencies, with a wavelength so as to diffract electrons across the Fermi surface.  This can be achieved using Intrinsic Josephson Junction stacks based on BSCCO to generate THz electromagnetic radiation, coupled to a piezoelectric substrate to convert that to acoustic waves.  We suggest starting with simple superconductors such as Nb, and looking for substantial enhancements of $I_c$, $\Delta$, and $T_c$.  Further work may target cuprates and other high-temperature superconductors.

If these experiments can be demonstrated, they should provide a new technique for investigating new and improved superconducting materials and devices, as well as answering some fundamental questions as to the nature of superconductivity in these materials.

## References

[1] S.B. Kaplan, C.C. Chi, D.N. Langenberg, C.C. Chang, S. Jafarey, and D.J. Scalapino, "Quasiparticle and Phonon Lifetimes in Superconductors," *Phys. Rev. B.*, vol. 14, pp. 4854-4873, Dec. 1976.
[2] A.M. Kadin and A.M. Goldman, "Dynamical Effects in Nonequilibrium Superconductors: Some Experimental Perspectives", in *Nonequilibrium Superconductivity*, ed. by D.N. Langenberg and A.I. Larkin, Elsevier, Amsterdam, Netherlands, 1986, pp. 253-323.
[3] J.A. Pals, K. Weiss, P.M. van Attekum, R.E. Horstman, and J. Wolter, "Nonequilibrium Superconductivity in Homogeneous Thin Films,", *Phys. Reports,* vol. 89, no. 4, pp. 323-390, Sept. 1982.
[4] S.B. Kaplan, J.R. Kirtley, and D.N. Langenberg, "Experimental Determination of the Quasiparticle Energy Distribution in a Nonequilibrium Superconductor," *Phys. Rev. Lett.*, vol. 39, p. 291, Aug. 1977.
[5] R.G. Melton, J.L. Paterson, and S.B. Kaplan, "Superconducting Tunnel Junction Refrigerator", *Phys. Rev. B*, vol. 21, p. 1858, Mar. 1980.
[6] T.M. Klapwijk, J.N. van den Bergh, and J.E. Mooij, "Radiation stimulated superconductivity", *J. Low Temp. Phys*, vol. 26, p. 385, 1977.
[7] T.G. Tredwell and E.H. Jacobson, "Phonon-induced enhancement of the superconducting energy gap", *Phys. Rev. Lett.*, vol. 35, p. 244, 1975.
[8] M.G. Blamire, E.C. Kirk, J.E. Evetts, and T.M. Klapwijk, "Extreme critical temperature enhancement of Al by tunneling in Nb/AlOx/Al/AlOx/Nb tunnel junctions." *Phys. Rev. Lett.*, vol. 66, p. 220, Jan. 1991.

[9] P.J. de Visser, et al., "Nonequilibrium response of a superconductor to pair-breaking radiation measured over a broad frequency band," *Appl. Phys. Lett.*, vol. 106, art. 252602, June 2015.

[10] F. Laviano, et al., "THz detection above 77 K in YBCO films patterned by heavy-ion lithography," *IEEE Sensors Journal*, vol. 10, no. 4, pp. 863-868, Apr. 2010.

[11] J. Demsar, "Light-Induced Superconductivity," *Nature Physics*, vol. 12, pp. 202-203, March 2016.

[12] M. Beck, et al., "Probing Superconducting Gap Dynamics with THz Pulses," *Proc. CLEO 2015*, paper SM3H.3, May 2015. DOI: 10.1364/CLEO_SI.2015.SM3H.3

[13] G.L. Dakovski, et al., "Enhanced coherent oscillations in the superconducting state of underdoped $YBa_2Cu_3O_{6+x}$ induced via ultrafast terahertz excitation," *Phys. Rev. B*, vol. 91, art. 220506(R), June 2015.

[14] A. Patz, et al., "Ultrafast THz probes of nonequilibrium Cooper pairs in iron pnictides", *Proc. CLEO: QELS Fundamental Science 2016*, paper FTu3L.2, June 2016. DOI: 10.1364/CLEO_QELS.2016.FTu3L.2

[15] W. Li, C. Zhang, X. Wang, J. Chakhalian, and M. Xiao, "Ultrafast spectroscopy of quasiparticle dynamics in cuprate superconductors," *J. Magnetism and Magnetic Materials*, vol. 376, pp. 29-39, Feb. 2015.

[16] G. Grüner, *Density Waves in Solids*, Addison Wesley, 1994; reprinted by Perseus Books, 2000.

[17] G. Grüner and A. Zettl, "Charge Density Wave Conduction: A Novel Collective Transport Phenomenon in Solids", *Phys. Reports*, vol. 119, no. 3, pp. 117-232, 1985.

[18] H.Fröhlich, "On the theory of superconductivity: one-dimensional case", *Proc. R. Soc.*, vol. A223, p. 296, 1954.

[19] W. Kohn, "Image of the Fermi surface on the vibration spectrum of a metal", *Phys. Rev. Lett.*, vol. 2, p. 393, 1959.

[20] A.M. Kadin, "Coherent Lattice Vibrations in Superconductors", *Physica C: Supercond.*, vol. 468, no. 4, pp. 255-259, Feb. 2008.

[21] A.M. Kadin, "Superconductivity Without Pairing", *ArXiv Physics Preprint*, 2009. http://arxiv.org/abs/0909.2901

[22] S.B. Kaplan and A.M. Kadin, "Superconductivity via Two-Phase Condensation of Localized Electrons", *APS March Meeting* (2012). http://meetings.aps.org/Meeting/MAR12/Event/162302

[23] A.M. Kadin, "Josephson Junctions Without Pairing?, *ArXiv Physics Preprint,* 2010. http://arxiv.org/abs/1007.5340

[24] P. Aynajian, et al., "Energy gaps and Kohn anomalies in elemental superconductors", *Science,* vol. 319**,** p. 1509 (2008).

[25] D.J. Scalapino, "This Coincidence Cannot Be Accidental", *Science*, vol. 319, p. 1492 (2008).

[26] S. Johnston, et al., "Coincidence between Energy Gaps and Kohn Anomalies in Conventional Superconductors," *Phys. Rev. B*, vol. 84, art. 174523, Nov. 2011.

[27] R. Chaudhury and M.P. Das, "Kohn Singularity and Kohn Anomaly in Conventional Superconductors: Role of Pairing Mechanism", *J. Physics: Condensed Matter*, vol. 25, no. 12, art. 122202, Feb. 2013.

[28] D.H. Torchinsky, et al., "Fluctuating charge density waves in a cuprate superconductor," *Nature Materials*, vol. 12, pp. 387-391, 2013.

[29] W.D. Wise, et al., "Charge density wave origin of cuprate checkerboard visualized by scanning tunneling microscopy," *Nature Physics*, vol. 4, pp. 696-699, 2008.

[30] M. Le Tacon et al., "Intense paramagnon excitations in large family of high-temperature superconductors," *Nature Physics*, vol. 7, pp. 725-730, Sept. 2011.

[31] R. Mankowsky, et al., "Nonlinear lattice dynamics as a basis for enhanced superconductivity in YBCO," *Nature*, vol 516, p. 71, Dec. 2014.

[32] Z.M. Raines, et al., "Enhancement of superconductivity via periodic modulation in a 3D model of cuprates," *Phys. Rev. B,* vol. 91, art. 184506, May 2015.

[33] M.A. Sentef, et al., "Theory of light-enhanced phonon-mediated superconductivity," *Phys. Rev. B*, vol. 93, 144506, Apr. 2016.

[34] A. Komnik and M. Thorwart, "BCS theory of driven superconductivity," *ArXiv Physics Preprint*, 2016. https://arxiv.org/abs/1607.03858 .

[35] D.A. Williams, "Coherent Nonequilibrium Phonons: An Induced Peierls Distortion," *Phys. Rev. Lett.*, vol. 69, no. 17, p. 2551, Oct. 1992.

[36] K.W. Kim, et al., "Ultrafast transient generation of spin-density-wave order in the normal state of $BaFe_2As_2$ driven by coherent lattice generation," *Nature Materials*, vol. 11, pp. 497-501, Apr. 2012.

[37] A. Singer, et al., "Photoinduced enhancement of the charge-density-wave amplitude," *Phys. Rev. Lett.*, vol. 117, 056401, May 2015.

[38] W. Grill and O. Weis, "Excitation of coherent and incoherent terahertz phonon pulses in quartz using infrared laser radiation," *Phys. Rev. Lett.*, vol. 35, p. 588, Sept. 1975.

[39] W.E. Bron, M. Rossellini, Y.H. Bai, and F. Keilmann, "Surface requirements for piezoelectric generation of high-frequency phonons," *Phys. Rev. B*, vol. 27, p. 1370, Jan. 1983.

[40] I. Kakeya and H. Wang, "THz emission from Bi2212 intrinsic Josephson junctions", *Supercond. Sci. Technol*. 29, 073001, 2016.

[41] T. Kashiwagi et al., "High-$T_c$ intrinsic Josephson junction emitter tunable from 0.5-2.4 THz", *Appl. Phys. Lett*. 107, 082601, 2015.

[42] U. Welp, K. Kadowaki, and R. Kleiner, "Superconducting emitters of THz radiation", *Nature Photonics*, vol. 7, pp. 702-710, Aug. 2013.

[43] D.Y. An, et al., "THz emission and detection both based on high-$T_c$ superconductors: Toward an integrated receiver", *Appl. Phys. Lett*. vol. 102, 092601, 2013.

[44] S. Cherepov, et al., "Electric-field-induced spin wave generation using multiferroic magnetoelectric cells," *Appl. Phys. Lett*, vol. 104, 082403, 2014.

[45] M.R. Armstrong, et al., "Observation of THz radiation coherently generated by acoustic waves", *Nature Physics*, vol. 5, 285-288, 2009.

[46] D.E. Grupp and A.M. Goldman, "Giant piezoelectric effect in $SrTiO_3$ at cryogenic temperatures", *Science*, vol. 276, p. 392, 1997.